\documentclass[10pt,journal,compsoc]{IEEEtran}
\ifCLASSOPTIONcompsoc
  \usepackage[nocompress]{cite}
\else
  \usepackage{cite}
\fi
\ifCLASSINFOpdf
\else
\fi
\usepackage{amsmath}
\usepackage{graphicx}
\usepackage{subfigure}
\usepackage{algorithm}
\usepackage{algpseudocode}
\usepackage{float}
\usepackage{setspace}
\usepackage{array}

\begin{document}
%
\title{Secure and Efficiently Searchable IoT Communication Data Management Model: Using Blockchain as a new tool}

\author{Ziqing~Guo,~\IEEEmembership{Student~Member,~IEEE,}
        Hua~Zhang,~\IEEEmembership{Member,~IEEE,}
        Zhengping~Jin,
        and~Qiaoyan~Wen
\thanks{All authors of this paper are with State key Laboratory of Networking and Switching Technology, Beijing University of Posts and Telecommunications, Beijing 100876, China. Corresponding author: Hua Zhang, e-mail: zhanghua\_288@bupt.edu.cn.}}

\IEEEtitleabstractindextext{%
\begin{abstract}
With the rapid development of the Internet of things (IoT), more and more IoT devices are connected and communicate frequently. In this background, the traditional centralized security architecture of IoT will be limited in terms of data storage space, data reliability, scalability, operating costs and liability judgment. In this paper, we propose an new key information storage framework based on a small distributed database generated by blockchain technology and cloud storage. Specifically, all encrypted key communication data will be upload to public could server for enough storage, but the abstracts of these data (called "communication logs") will be recorded in "IoT ledger" (i.e., an distributed database) that maintained by all IoT devices according to the blockchain generation approach, which could solve the problem of data reliability, scalability and liability judgment. Besides, in order to efficiently search communication logs and not reveal any sensitive information of communication data, we design the secure search scheme for our "IoT ledger", which exploits the Asymmetric Scalar-product Preserving Encryption (ASPE) approach to guarantee the data security, and exploits the 2-layers index which is tailor-made for blockchain database to improve the search efficiency. Security analysis and experiments on synthetic dataset show that our schemes are secure and efficient.
\end{abstract}
}

\maketitle
\IEEEdisplaynontitleabstractindextext

%
\IEEEpeerreviewmaketitle
\IEEEraisesectionheading{\section{Introduction}\label{sec:introduction}}

\subsection{Motivation}

\IEEEPARstart{T}{he} Internet of things (IoT) is developing toward high intelligence and detailed division of labor \cite{J2016Bitcoin, Wang2016Practical}. It is common that each IoT device in the same network only need to deal with their "own job", while the operation of whole task will be achieved by data sharing or instruction switching among these IoT devices. However, this development tendency brings great convenience as well as new problem for IoT usage. Specifically, the behaviour of one device may be the result of multiple related devices' actions. In industrial control systems that with high security requirements, once there are some accidents happened, finding the wrong links according to IoT devices' key communication data will naturelly becomes the user's first task.

\begin{figure}
\centering
\includegraphics[width=3in]{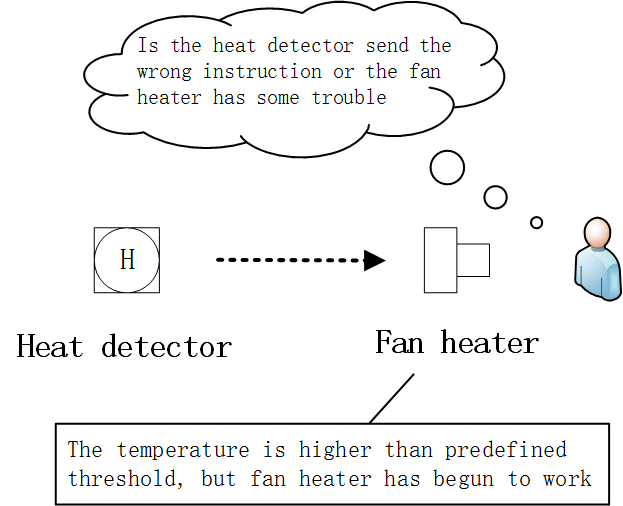}
\caption{\footnotesize{An accident that may occured in IoT}}
\label{scenario}
\end{figure}

An example can be shown as Fig.\ref{scenario}. A heat detector and a fan heater work together to control the temperature. One day user find that the fan herater has worked when the temperature is higher than predefined threshold. Obviously, there may be two reasons for this accident: the heat detector sent the wrong temperature data to the fan heater, or the fan heater received the correct information but there were something wrong with itself. To solve this problem, the straightford way is examining the communication data between they two. Therefore, it is necessary to design a trusted, secure and high-performance IoT communication data management system.

In fact, it is not difficult for the traditonal data management architecture to provide communication data management system. There are 4 straightforward frameworks that could maintain the IoT communication data and enable user to search data he/she interested under the help of IoT devices, but all these 4 frameworks have some limitions regarding to efficiency and security. We specifically analyze them as follows.

\subsubsection{IoT without Center}
In this framework, there is no center server responsible to monitor and store the communication data of entire network, so that each IoT device has to maintain its own communication data on local site. Therefore the following three limits are inevitable. Assuming the IoT device $R$ interacts with other $\alpha$ devices $\{S_1, S_2, \dots, S_{\alpha}\}$.

\begin{itemize}
  \item Constrained Storage: It is difficult for some lightweight storage IoT devices to store all related communication data which increases over time.

  \item Poor Data Reliability: Since all communication data for device $R$ are stored by itself, there is a potential risk that the data will be tampered or lost if $R$ malfunctions or suffers external attacks, moreover, once it happens, recovering the original data is difficult.

  \item Poor Scalability: Both the bandwidth and storage space of one device $R$ would be challenged when the number of IoT devices that $R$ linked increases.
\end{itemize}

\subsubsection{IoT with One Local Center}
In this framework, there is only one center server responsible to monitor and store the communication data of entire network, so that all IoT devices do not need to store anything, which would solve the problem of constrained storage for lightweight IoT devices, but it still faces the following two limits.

\begin{itemize}
  \item Poor Data Reliability: Due to the potential risk that the center server may suffer the single point of failure, once the data it stored are tampered or lost, it is difficult to recover the original data.

  \item Poor Scalability: Like the model "IoT without Center", it is difficult to extend the scale of IoT under "one center" framework.
\end{itemize}

\subsubsection{IoT with Multiple Local Centers}
In this framework, there are multiple center servers responsible to monitor and store the communication data of entire network, and each of them maintains one duplicate of all communication data. Compared with "IoT with One Center", this model solves the poor data reliability problem produced by center server's single point of failure, and also could provide enough storage. However, it still faces the following two limits.

\begin{itemize}

  \item Poor Scalability: Since each center server maintains one duplicate, the storage burdren of one center server does not mitigate. It is also difficult to extend the scale of IoT under "multiple centers" framework.

  \item High Operating Costs: The increased number of center servers will increase the operating costs of IoT, especially for some small application scenarios (such as smart home), the increased operating costs can not be accepted.
\end{itemize}

\subsubsection{IoT with Cloud}

The combination between IoT and cloud computing has become a popular research topic in recent years. In IoT application, due to the unconstrained computation and communication resources for cloud computing, the cloud server could play the role of center server, which is responsible to monitor, process and store the IoT data. As for IoT communication data management, a straightforward solution is all IoT devices upload their encrypted (for security purpose) communication data to cloud server and user could query on these data in cloud platform. Compared with the model "IoT with Multiple Centers", exploiting the cloud computing to IoT could solve the problems of poor scalability and high cost, but it still exist two limits.

\begin{itemize}
  \item Poor Data Reliability: Like the model "IoT with One Center", once the data cloud stored are tampered or lost, it is difficult to recover the original data.

  \item Difficult Liability Judgment: Unlike the model "IoT with One Center", the cloud server can not be full controlled by user, so that if the cloud server lose the user's data (due to the internal faults or external attacks), it may deny that it received the data, and user can not determine whether it lies.
\end{itemize}

In summary, all above 4 IoT communication models can not simultaneously achieve the properties of (1) enough storage, (2) good data reliability and (3) reliable liability judgment, (4) scalability and (5) cheap operating costs. So it is necessary to design a new ICDM model that satisfy all these requirements. Besides, to facilitate user efficiently search on IoT communication data and not reveal anything privacy, a secure and efficient search scheme for the ICDM model is also of critical importance.

\subsection{Our Contributions}

In this paper, we propose a novel ICDM model based on blockchain technology which could achieve better performance in terms of the 5 aspects we mentioned in above paragraph. Specifically, all IoT devices are regarded as nodes in blockchain network, they outsource the encrypted commmunication data to public cloud and then broadcast their encrypted communication logs (like the digest communication data) as well as corresponding index to entire network. After that, all nodes will cooperate in generating new block based on broadcasted information, and the complete blockchain is maintained by some IoT devices with relatively enough storage space. From the above, we can see that the cloud storage provide enough storage space for communication data; the small size communication logs contained in blockchain and the tamper-resistant property of blockchain guarantees data reliability and reliable liability judgment; the decentralized management framework make the scheme would not constrained by the growth of network scale. Besides, for supporting secure and efficiently search on IoT communication logs, we construct encrypted index for searching on IoT communication logs which stored in blockchain: the ASPE algorithm \cite{Wong2009Secure} is exploited for carrying out range query on encrypted communication logs and not reveal any privacy; the two layers index (time series index and kd-tree index) is exploited for improving search efficiency. In conclusion, our major contributions can be summarized as follows.

\begin{itemize}
  \item For the first time, we propose a practical IoT communication data management (ICDM) model based on blockchain. Compared with the existing 4 ICDM models we mentioned above, our model could achieve better performance in storage capacity, data reliability, scalability and operating costs simultaneously.

  \item We design a novel 2 layers index for efficiently searching on IoT communication logs. Furthermore, the combination of Merkle tree and kd tree we exploit in guarantees the tamper-resistant of the 2nd layer index.

  \item Thorough security analysis and experiments on synthetic data show our proposed scheme is secure and supports efficient search on IoT communication log.
\end{itemize}

The reminder of this paper is organized as follows. Section \ref{Overview} introduces the notations and some related preliminaries, proposes the system and threat models, summarizes the design goals of our blockchain-based ICDM model. Followed by Section \ref{BB_ICDM} and Section \ref{BB-Search}, we describes the principal of our blockchain-based ICDM model and the scheme for searching IoT communication log, respectively. The security and performance analysis are presented in Section \ref{SecPerAnalysis}, and the related works are discussed in Section \ref{RelatedWork}. Finally, Section \ref{Conclusion} covers the conclusion.

\section{Overview}\label{Overview}

\subsection{Notations}

\begin{itemize}
\item $\mathcal{D}$ - The IoT device collection, which is composed by $n$ IoT devices $\mathcal{D} = \{D_1, D_2, \dots, D_n\}$.

\item $\mathcal{M}$ - The miners collection, which is the subset of $\mathcal{D}$. The miners are unoccupied IoT devices who would like to generate the new block for broadcasted information.

\item $\mathcal{F}_i$ - The communication data files, denoted as a collection of $m_i$ communication data files that maintained by the IoT device $D_i$. $\mathcal{F}_i = \{F_{i, 1}, F_{i, 2}, \dots, F_{i, m_i}\}$.

\item $\mathcal{L}_i$ - The communication log collection of the IoT device $D_i$, which is composed by $m_i$ formalized communication data files (called communication logs), denoted as $L_i = \{L_{i, 1}, L_{i, 2}, \dots, L_{i, m_i}\}$.

\item $\tau_i$ - The plaintext form of kd-tree for the communication log collection $L_i$.

\item $\tau_i^*$ - The secure searchable IMT index that is encrypted from $\tau$.

\item $B$ - The blockchain, which is composed by a series of ordered blocks: $B_1, B_2, \dots, B_{cur}$, where $B_j$ denotes the $j$-th block, and $B_{cur}$ denotes the last block in $B$ currently.

\item $\mathcal{Q}$ - The plaintext form of user's range query (i.e., a $l$-dimensional hyperrectangle), denoted as a collection composed by $2l$ pairs of anchor points $Q = \{A_1^{in}, A_1^{out}; A_2^{in}, A_2^{out}; \dots; A_{2l}^{in}, A_{2l}^{out}\}$.

\item $TR$ - The ciphertext form of query $\mathcal{Q}$, which plays as a trapdoor for search request.

\end{itemize}

\subsection{Preliminaries}

\subsubsection{Blockchain}
The blockchain has become a popular technology due to its good properties like decentration, tamper-resistant and distributed accounting.

Decentralized cryptocurrency (e.g., Bit- coin \cite{J2016Bitcoin}) has gained popularity and is also quoted as a glimpse in the future . The cryptocurrency system builds on top of a novel technology named blockchain , which is essentially a distributed database of transactions. Digital information has been executed and shared among participating parties and allows public ledger of all transactions. A blockchain is composed of verifiable records for each single transaction ever made which is verified by consensus of a majority of the participants in the system. Blockchain technology is finding applications in wide range of non-financial areas besides current financial areas, such as decen- tralized proof of the existence of documents , decentralized IoT  and decentralized storage .
Blockchain-based storage has become a newly of growth engine in data sharing since it does not need a central service provider. We find that the data retrieval approaches are rarely studied in the blockchain-based system. To date, a personal data management system  was proposed with the assistance of the blockchain technology to ensure users own and control their data against the honest-but-curious services. A decentralized smart contract system named Hawk was proposed to retain transaction privacy from the public’s view, while no detailed retrieval algorithm was given. A healthcare chain was constructed to facilitate data interoperability in health information networks. However, these systems focus on the concepts with the corresponding frameworks instead of the concrete algorithms to guarantee data utilization and data secrecy. Also, the linkable transaction privacy is still a margin in the blockchain-based retrieval.

\subsubsection{Merkle Tree}

A merkle tree(hash tree) is a tree in which every leaf node recods the hash of a data block and every non-leaf node stores the hash of its child nodes' values. Most merkle trees are binary, in which every node has two child nodes\cite{Merkle1987A}. As shown in Fig.\ref{index2}, for leaf nodes with labels $h_1,h_2$, their father node's label is the hashed of his children such as $h_{12}=h(h_1||h_2)$, where $||$ means concatenation. Merkle tree is used to effectively verify data stored  or transferred between computers, specially, to verify whether there exsits a data record in block in BC, the computation cost of demonstrating a leaf node is a part of a given binary hash tree is no more than $log(N)$(N is the total number of leaf nodes); this contrasts with hash lists, where the cost is proportional to the number of leaf nodes itself\cite{Lee1977Worst}.

\subsubsection{Secure Range Query (SRQ)}

The secure range query (SRQ) has been a hot research topic recent years \cite{Wang2016Practical}. Its purpose is to search on an encryptd database with an encrypted range query (such as an encrypted hyperrectangle or hypersphere), and finally outputs the encryptd points who are in this range. In our search scheme, after locking the target blocks that needs to be further searched, user would like to carry out SRQ on our 2nd index (i.e., encrypted Indexed-Merkle tree) with rectangular query for searching interested communication logs. The SRQ algorithm we exploit in this paper is based on the "Halfspace Range Query" approach proposed Wang et.al. \cite{Wang2013Secure}, Here we briefly introduce how it works.

Firstly we introduce how to determine the location relationship between a point and a hyperplane. As given in \cite{Wang2013Secure}, there is a hyperplane $H$, which could divide the space into 2 halfspaces, denoted as $H^1$ and $H^2$. Now, we can give 2 anchor points $A^1 and A^2$ for $H$, where the segment $A^1A^2$ is normal to $H$ and equidistance from $H$. We can find that, for any point $P$ in this $l$-dimensional space, if $dist(A^1, P) > dist(A^2, P)$, $P$ is in the halfspace $H^1$; if $dist(A^1, P) < dist(A^2, P)$, $P$ is in the halfspace $H^2$; and if $dist(A^1, P) = dist(A^2, P)$, $P$ is on the $H$, where $dist()$ denotes the Euclidean distance of two points. Therefore, we can determine $P$ is in $H^1, H^2$ or on the $H$ by computing the distance between $A^j, j \in \{1, 2\}$ and $P$.

Secondly we show how to determine whether a point is in an hyperrectangle under $l$-dimensional space. Since there are $2l$ hyperplanes for a $l$-dimensional hyperrectangle $Q$, we need to totally prepare $4l$ anchor points. For each hyperplane $H_i$, the halfspace contains $R$ is denoted as $H_i^{in}$, and the other halfspace is denoted as $H_i^{out}$. For point $P$, if and only if $P$ is in all $H_i^{in}$, $P$ is in $Q$  Therefore, we can determine whether $P$ is in $Q$ by computing the the distance between $P$ and each pair of anchor points $A_i^{in}$ and $A_i^{out}$, shown in Algorithm \ref{isPointInRect}.

\begin{algorithm}
  \caption{\emph{isPointInRect($P$, $R$)}}\label{isPointInRect}
  \begin{algorithmic}[1]
  \Require the $l$-dimensional point $P$ and the hyperrectangle $Q$ which is indicated by a $4l$ anchor points collection, $Q = \{A_1^{in}, A_1^{out}; A_2^{in}, A_2^{out}; \dots; A_{2l}^{in}, A_{2l}^{out}\}$
  \Ensure a boolean value that denotes whether $P$ is in $Q$
  \For{each $A_i^{in}$ and $A_i^{out}$ in $Q$}
    \If{$dist(P, A_i^{in}) > dist(P, A_i^{out})$}
      \Return false
    \EndIf
  \EndFor\\
  \Return true
  \end{algorithmic}
\end{algorithm}

Thirdly we show how to determine whether a hyperrectangle $Q$ interact with another hyperrectangle $R$. According to the theorem proposed by Wang et.al., if the 2 extremal vertices (denoted as $V_1, V_2$) of $R$ are both not in any $H_i^{in}$ of $Q$, $Q$ would not interact with $R$, where the extremal vertices are vertices of $R$ who has the maximum and minimum coordinates (like lower-left vertex and upper-right vertex in 2-dimensional rectangle). The detail is shown in Algorithm \ref{areRectsInter}.

\begin{algorithm}
  \caption{\emph{areRectsInter($Q$, $R$)}}\label{areRectsInter}
  \begin{algorithmic}[1]
  \Require the hyperrectangle $Q$ which is composed by a $4l$ anchor points $Q = \{A_1^{in}, A_1^{out}; A_2^{in}, A_2^{out}; \dots; A_{2l}^{in}, A_{2l}^{out}\}$, and the hyperrectangle $R$ which is composed by 2 extremal vertices $R = \{V_\bot, V_\top\}$.
  \Ensure a boolean value that denotes whether $Q$ is interact with $R$
  \For{each $A_i^{in}$ and $A_i^{out}$ in $Q$}
    \If{$dist(V_\bot, A_i^{in}) > dist(V_\bot, A_i^{out})$ and $dist(V_\top, A_i^{in}) > dist(V_\top, A_i^{out})$}\\
      \Return false
    \EndIf
  \EndFor\\
  \Return true
  \end{algorithmic}
\end{algorithm}

In above two algorithms, the key computational steps are Euclidean distance of two $l$-dimensional points, and such computation can be efficiently realized by the Asymmetric Scalar-product Preserving Encryption (ASPE) \cite{Wong2009Secure} approach in encrypted environment. Hence, the Algorithm \ref{isPointInRect},\ref{areRectsInter} combined with ASPE is our main technological method when user searches encrypted communication logs stored in blockchain.

\subsection{System Model}\label{Sysmodel}

\begin{figure}[H]
\centering
\includegraphics[width=3in]{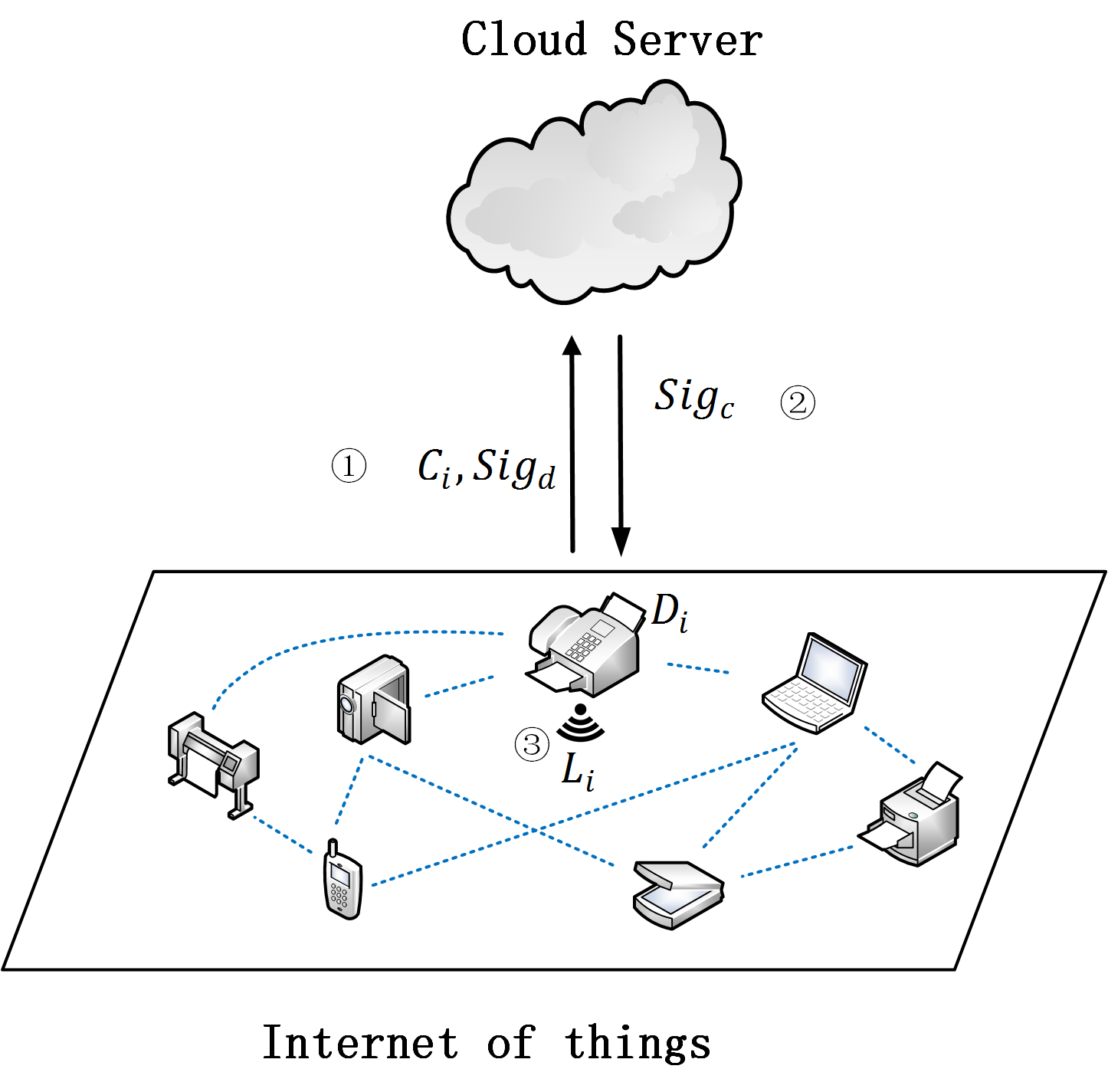}
\caption{\footnotesize{The architecture of privacy-preserving LRS service}}
\label{model}
\end{figure}

As shown in Fig.\ref{model}, the system model of our scheme involves three kinds of entities: IoT devices, cloud server and user, where some unoccupied IoT devices will play the role of miners. We now describe these entities in detail as follows.

\textbf{\emph{IoT Devices}} is a collection composed by all IoT devices. During the the process of IoT system running, an IoT device would communicate with other related devices for switching orders or sharing data. Originally, each IoT device should maintain its own communication in case user inquires it. But with the limit of storage capacity, not all IoT devices have enough space to store their whole communication data that grow over time. In our scheme, once the cumulative size of maintained communicate data exceeds the storage upper bound of one IoT device (denoted as $D_i$), $D_i$ would first upload the integrated ciphertext form of communication data into cloud server, and then broadcast the formalized communication data (denoted as communication logs $L_i$, including the hash values of communication data, the corresponding signatures and the index used to search) to entire network. After that, the miners (shown in next paragraph) construct a blockchain for recording these communication logs, which will support user's search task. Finally, $D_i$ could empty its local storage. Note that, in general, different IoT devices have different storage capacities, so the number of broadcasted communication logs from different IoT devices are different, too.

\textbf{\emph{Miners}} is a subset of whole IoT devices. In fact, the miners are unoccupied IoT devices that are not busy dealing their own business when an IoT device $D_i$ broadcasts its formalized communication log collection $L_i$. In our scheme, once a miner receives $L_i$, it will generate a new block $B$ that contains $L_i$ according to the predefined difficulty, and then publish this new block. If there are more than half of IoT devices in the entire network approve the validity of block $B$, $B$ can join up to the main blockchain. Note that the \emph{Miners} is a dynamic collection, its members are only decided by who are unoccupied when there are IoT devices broadcast their formalized communication logs. For convenience, we denote all IoT devices except miners are "Busy IoT Devices".

\textbf{\emph{Cloud Server}} is responsible for storing the integrated communication logs. In our scheme, IoT devices will upload all of its encrypted communication logs into cloud platform before broadcasting the logs' digests and corresponding signatures to entire network. So that user could retrieve the integrated encrypted communication logs from cloud server according to the corresponding storage indexes contained in blockchain. Finally, user decrypts the encrypted communication logs and obtains integrated communication logs.

\textbf{\emph{User}} searches on blockchain by some data attributes, in terms of communication data, its timestamp, keywords, corresponding sender and receiver are all can be regarded as effective search items. The properties of distributed storage and tamper-resistant guarantee that user could obtain the true and complete formalized communication logs. Moreover, if user would like to retrieve integrated communication data, he/she could retrieve it from cloud server by corresponding storage indexes contained in blockchain. Note that all search processes are executed on ciphertext.

\subsection{Threat Model}\label{Thrmodel}

The IoT devices and cloud server in our scheme are both "honest but curious and weak". Specifically,

As for \textbf{\emph{IoT Devices}}, no matter they play the role of miner or busy IoT devices, they would honestly and correctly executes instructions in the designated protocol, but they are curious to infer and analyze the received data, including the content of other IoT devices' formalized communication logs and the users' queries. Furthermore, due to some internal and external factors (like equipment failures or external attack), a block stored in some IoT devices also possibly be corrupted (damaged or even lost), that is why we call IoT devices are "weak".

As for \textbf{\emph{Cloud Server}}, it would honestly execute the protocol, but it is curious about the plaintext content of the uploaded communication data. Besides, if the cloud data is damaged or even lost, maybe the cloud server would shift the duties onto others. For example, it says that it has never received this piece of data.

\subsection{Design Goals}

Our proposed blockchain-based IoT communication data management (ICDM) model should achieve the following design goals simultaneously.

\begin{itemize}
\item \textbf{Tamper-resistant Communication Log}: To protect the authenticity and completeness of IoT communication logs, our proposed ICDM model should guarantee that any malicious participants can not tamper any maintained communication logs as soon as there are more than half of IoT devices honestly execute the protocol.

\item \textbf{Data Retrievability}: For communication logs, to prevent data loss risk due to the IoT device's single point of failure, the ICDM model should enable the user to retrieve any communication logs, unless all IoT devices who store the blockchain data miss the data; for communication data, if the cloud server thoroughly loss the data, the ICDM model enable user to obtain corresponding communication summary information by communication logs.

\item \textbf{Enough Storage}: With the rapid increase of connected IoT devices amount, our ICDM model should guarantee that even the communication data (logs) of lightweight storage IoT devices could be maintained completely.

\item \textbf{Efficient and Secure Range Query}: The proposed ICDM model aims to achieve efficient secure range query for communication logs on IoT devices. In the search process, the index privacy and query privacy should not be revealed to IoT devices, and our scheme also should guarantee that the communication logs' indexes are tamper-resistant.
\end{itemize}

For the last point we mentioned above, the index privacy and query privacy in our scheme are extremely similar to previous secure range query schemes \cite{Wang2013Secure}, so we do not describe them here in detail. In Section \ref{SecPerAnalysis}, we will analyze how our ICDM model achieves these privacy requirements.

\section{The proposed schemes}\label{ProposedSchemes}

In this section, we first show the framework of the novel blockchain-based ICDM model, where the specific working steps of different roles will be explained in detail. And then we design the secure and efficient data search scheme on encrypted IoT communication logs stored in blockchain, including the construction of two layers index and corresponding range query algorithm.

\subsection{Blockchain-based ICDM model}\label{BB_ICDM}

In Section \ref{Sysmodel}, we have briefly introduced the roles of 4 different entities, now we describe a step-by-step scheme to show their specific work.

\subsubsection{Upload communication data}

As we mentioned in Section \ref{Sysmodel}, in certain period, one IoT device $D_i$ generally communication with not only one IoT device, and each of these IoT devices will switch with $D_i$ for not only one piece of message. Therefore, the communication data file collection $F_i$ can be formally defined as follows.
\[
F_i = \{(F_{i1}^{\alpha}, F_{i2}^{\alpha}), (F_{i3}^{\beta}, F_{i4}^{\beta}, F_{i5}^{\beta}), \dots, (F_{i(m-1)}^{\gamma}, F_{im}^{\gamma})\}
\]

Where $F_{it}^{j}, t \in \{1, \dots, m\}$ denotes one communication data file between $D_i$ and $D_{j}$.

We assume there are totally $n$ devices communicate with $D_i$, and $D_i$ totally stores $m$ data files in this period. Note that $m$ is depend on the size of each $F_{i_t}^{j}$ and the storage capacity of $D_i$, i.e., size $m$ is different for different IoT devices, even different for the same device in different periods.

When the cumulative size of communication data maintained by $D_i$ has exceeded its storage upper bound, $D_i$ will upload the ciphertext form of $F_i$ (denoted as $C_i = \{C_{i1}, C_{i2}, \dots, C_{im}\}$) to public cloud server. For confirming this upload process, $D_i$ and the cloud server need to generate corresponding signature as below. After verifying the validity of signateurs, they would mutual hold each other's signature.

\begin{itemize}
\item $Sig_c(h_1||h_2||\dots||h_m||ID(D_i)||TS)$: Signed by cloud server, where $h_t$ is the hash value of $t$th encrypted data files in $C_i$, $ID(D_i)$ is the identification of $D_i$, $TS$ denotes the timestamp of upload process, and the symbol $||$ means the cascade. This signature is briefly denoted as $Sig_c$ and hold by $D_i$.

\item $Sig_d(h_1||h_2||\dots||h_m||ID(C)||TS)$: Signed by $D_i$, where $ID(C)$ denotes the identification of cloud server. This signature is briefly denoted as $Sig_d$ and hold by cloud server.
\end{itemize}

\subsubsection{Generate communication logs}\label{ComLogs}

After uploading encrypted data to cloud server, the IoT device $D_i$ need to generate communication log collection $L_i$ for $F_i$ and broadcast it to entire network, so that the miners could record it in new block. In fact, $L_i$ is similar to the transactions in bitcoin network, it should contain the digest of communication data and corresponding authentication information. Specifically, for any communication file $F_{it}^{j}$, no matter it is the information that $D_i$ send to $D_{j}$ or $D_{j}$ send to $D_i$, to confirm the authenticity of this communication behavior, the signatures of both communication parties are necessary. In summary, the communication log collection $L_i$ can be formly described as follows.

\[\begin{split}
L_i =& \{(E(h(F_{i1})), sig_{\alpha}), (E(h(F_{i2})), sig_{\alpha}),(E(h(F_{i3})), sig_{\beta}),\\
&\dots, (E(h(F_{im})), sig_{\gamma})\}=\{L_{i1}, L_{i2}, L_{i3}, \dots, L_{im}\}
\end{split}\]

Communication log is composed by the following 4 items:

\begin{itemize}
\item $E(h(F_{it}))$: The encrypted hash value of communication data file $F_{it}^{j}$.

\item $sig_{j}$: Signed by $D_{j}$, it means $sig_{j}((E(h(F_{it})||TS)$, where $F_{it}$ is sent from $D_j$ to $D_i$.

\item $Sig_c(h_1||h_2||\dots||h_m||ID(D_i)||TS)$: The signature of cloud server, it is same to the above in section几. This signature is briefly denoted as $Sig_c$ hereafter.

\item $Sig_d$: Signed by the communication data owner $D_i$, it means $Sig_d(h(L_i||Sig_c))$

\end{itemize}

Finally, the whole communication logs can be formly written as $L_i||Sig_c||Sig_d(h(L_i||Sig_c))$. And it is broadcasted by $D_i$ to entire network.

\subsubsection{Generate new block}

\begin{figure}[H]
\centering
\includegraphics[width=3in]{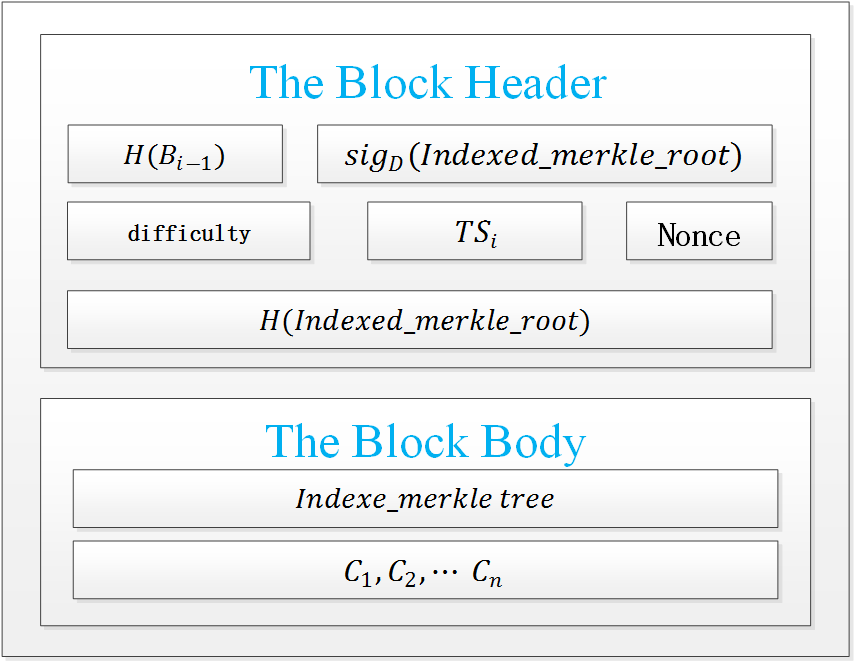}
\caption{\footnotesize{The detials of a new block}}
\label{block}
\end{figure}

Bitcoin network uses a proof-of-work(PoW) system to generate new block. This work is often called bitcoin mining. As same as bitcoin, in our scheme, new block is generated by proofing of work. The detials of a new block is shown in Fig.\ref{block}. A new block is composed of block header and block body. In the block header, $H(B_{i-1})$ is hash value of father block, $sig_D(Indexed\_merkle\_root)$ is device's signature of root value in constructed indexed merkle tree(IMT), difficulty is predefined difficulity target of PoW algorithm before constructing this block, $TS_i$ represents time stamp, Nonce is a counter used in PoW algorithm. In the block body, beside the communication logs $L_{i1}, L_{i2}, L_{i3}, \dots, L_{im}$ introducted in\ref{ComLogs}, there is a completely index tree $Indexed\_merkle tree$, whereas, it is the hash value of root in merkle tree in block used in bitcoin network. It is the key difference between block in our scheme and in bitcoin network, and is due to the requirement of subsequent index. Note that the detials of index are demonstrated in section\ref{BB-Search}.

There are some other differences in mining block between our proposed IoT network and traditional bitcoin network. In bitcoin network, selecting new legal block by adjusting difficulty of mining. To achieve better performance of practical operation, the difficulty predefined in our scheme is lower. One miner can get paid(bitcoin) after generating a legal block in bitcoin network.
Compared with incentive mechanism(paying bitcoin to miner) used in bitcoin model based BC, the propsed ICDM model incentes unoccupied IoT devices through other methods. For example, user can pay some money to the factory which manufactures or allocates the miner devices, or give score to devices or related factory. In addition, the incentive mechanism is out of scope of this paper, study of this field will be discussed in our future work.

Then the newly generated block is broadcasted to entire network. If it is approved by 51\% IoT devices, it can join up to the main chain.

\subsubsection{Analyses}
This section will give some analysis of constructing block in terms of storage, data reliability, reliable liability judgment, scalability and cheap operating costs as followings.

\begin{itemize}
\item \textbf{enough storage}: In traditional IoT based on cloud server, cloud server plays a role of maintaining the encrypted commmunication data, device only stores limited data. And in this scenario, cloud may deny it had received encrypted commmunication data when users wants to investigate in and assess the blame for the accident. To solve this, it is necessary for one device to record some information about its interation with other devices. In our proposed scheme, devices only needs to store main hash arrays of BC. And storage space is individually personalised depending on device's storage capacity and owner's preferences. Specifically, if one device's storage space is large enough, it can maintain the whole main chain in IoT network based BC. Otherwise, if one device's storage space is limited, it can only store short chain composed of the latest block. Every block contains a hash of the previous block. Given that property，correctness of all communication data are verifiable in the block chain, no matter how long the chain stored by device is.

\item\textbf{scalability}: With the extension of IoT network, data-storage burden dose not increase, because device can store chain within custom length. The designed decentral secure structure based on BC basically does not limite the scalability of IoT network. Because peer to peer network uses resources of all participating nodes and eliminates many-to-one traffic flows, of which the benefits is decreasing delay and overcoming the problem of a single point of failure.

\item\textbf{Data recover}: Beacause of formalization of communication data files, user can deduct approximate usable information stored in communication logs, even if the cloud data is damaged or lost.

\item\textbf{system operating costs}: In this paper, neither do we introdue additional center devices nor adopt double cloud server. Data transmission, such as broadcasting, and generating new block may cause communcation consumption and computational consumption, which is less compared with consumption of introducing a third party. As a whole, system operating costs in proposed scheme is more acceptable.

\end{itemize}

\subsubsection{Digitization of file content}
Cosider the content of every device's communication data file is including many ordered attributes, i.e., $F_{it}^{j}=\{attr_1, attr_2,\dots,attr_l\}$. For example, in former mentioned, the attributes of the heat detector's communication data file is temperate, wind keeps, wind direction and so on, namely $F_{heat\_detector}=\{temperate, wind\ power, wind\ direction\dots\}$. For every atrribute in communication data file, it relates attribute value. For example, $F_{heat\_detector}=\{temperate=26 degrees\ centigrade, wind\ power=strong, wind\ direction=south\dots\}$. All these contents can be be expressed by numbers, in this way, every file's content is mapped to a point, i.e., $F_{it}^{j}=\{attr_1, attr_2,\dots,attr_l\}$ can be represented as $P_{it}^{j}=(x_1, x_2,\dots,x_l)$. For example, $F_{heat\_detector}$ can be represented as $(26, 7, 3)$. We omit process of data normalization here.

\[\begin{split}
F_i = \{(F_{i1}^{\alpha}, F_{i2}^{\alpha}), (F_{i3}^{\beta}, F_{i4}^{\beta}, F_{i5}^{\beta}), \dots, (F_{i(m-1)}^{\gamma}, F_{im}^{\gamma})\}\\
\rightarrow P_i = \{(P_{i1}, P_{i2}, P_{i3}, P_{i4}, P_{i5}, \dots, P_{i(m-1)}, P_{im})\}
\end{split}\]
Finally, all the files belongs to one device can be digital as points in l-dimensional space area. Thereafter, every device constructes kd-tree for search.

\subsubsection{Construct kd-tree}\label{Constr_kdt}
A kd-tree can improve the search efficiency of range queries. The basic idea of building a kd-tree is to separate points by a hyperplane into two part at the same level and the different of the numbers of points in two part is no more than one, distribute them to a smaller bounding box in a higher level until there is only one point in one leaf node. In the proposed scheme, root node in kd-tree is a hyperrectangle $R\subset\mathcal{R}_l$ contians all the file points, is specified by its two extremal vertices $V_\bot=\{min\{x_1\}, \dots, min\{x_l\}\}$, $V_\top=\{max\{x_1\}, \dots, max\{x_l\}\}$.

\begin{algorithm}
  \caption{\emph{Construct kd-tree}}\label{constructe kd-tree}
  \begin{algorithmic}[1]
  \Require the communication data files $F_1, \dots, F_m$.
  \Ensure a kd-tree $\tau_i$
  \State Converting $F_1, \dots, F_m$ into point collection $u.S$ obtaining the inirial hyperrectangle $R$ that contains all $P_h\in u.S$
  \If{$S$ is empty}\\
    \Return None
  \Else
    \State Construct a empty tree$\tau_i = \phi$, with a root node $u$
    \If{$u.size=1$}
      \State Deleting point $P_h$ in $u$, $u.Pointers=\{P_c,P_l\}$ are point to corresponding $C_{it}$ and $L_{it}$
    \Else
       \State patitioning $u.R$ as 2 rectangles $R_1$,$R_2$ and $u.Pointers=\{P_1,P_2\}$ point to its child nodes $u_1$, $u_2$
       \For{$u_j\in\{u_1,u_2\}$}
         \State $node = node_j$, back to line 5
        \EndFor
    \EndIf
  \EndIf\\
  \Return$\tau_i$
 \end{algorithmic}
\end{algorithm}

\begin{itemize}\label{kd node}
\item $u$ -  a node of $\tau_i$ that has 4 attributes $\{ID, R, size, Pointers, S\}$, where $u.ID$ is the unique identifier of $u$; $u.R$ is the hyperrectangle$R(V_\bot,V_\top)$ represented by $u$; $u.size$ is the number of file points contained in $u.R$. If $u$ is leaf node, $u.Pointers$ consists of pointers that point to $C_{it}$ and to $L_{it}$, $t \in \{1, \dots, m\}$, whose corresponding $P_{it}$ in $u.R$. Otherwise, $u.Piointers$ consists of two pointers point to $u$'s two child nodes, respectively.$u.S$ is the set of all files points in $u.R$.

\item $GenID()$ - The function for generating the unique identifier of kd-tree node, $\{0,1\}^l\leftarrow GenID()$.
\end{itemize}
Notethat, hyperrectangle $R(V_\bot,V_\top)$ stored in one node$u$ is split into two small hyperrectangle $R_1(V_\bot,V_\top)$ and $R_2(V_\bot,V_\top)$ in child nodes by hyperplane $x_i'$ denotes as $X$, where $x_i$ is $ith$ dimensional median of all $P_h\in u.S$, $i$ is orderly selected dimension in $\{1, 2, \dots, l\}$. $u.S$ is split into two set by useing Algorithm\ref{isPointInRect} $isPointInRect(P_h,X)$. Algorithm\ref{constructe kd-tree} of constructing a kd-tree is shown below.

After constructing the plaintext index kd-tree, device deletes $u.S$ stored in every node, generats encryption of $u.R$ as $u.[R]=([V_\bot], [V_\top])$ by using ASPE, and replaces $R(V_\bot,V_\top)$ with $[R]=([V_\bot], [V_\top])$. Device uses its secret key $M$, an invertible $(l+1)\times (l+1)$ matrix. For $V = (v_1, v_2, \dots, v_l), V\in\{V_\bot,V_\top\}$, device creats $V_+ = V^T|1$ and obtains $[V] =M^{-1}V_+$. Finally, node $u$ has 4 attributes $\{ID, [R], size, Pointers\}$, the encrypted form of tree is denoted as $[\tau_i]$.

\subsubsection{Construct index merkle tree}
 After encrption of kd-tree, device construct an index merkle tree by bottom-up add context in construted kd-tree follows the algorithm\ref{constructe index merkle tree}.

\begin{itemize}
\item $u$ -  a node of $\tau_i^*$ that has 4 attributes $\{ID, [R], size, Pointers, hash\}$ The first three attributes are as same as defintion in \ref{kd node}. If $u$ is a leaf node, $u.hash=h([V_\bot]||[V_\top]||h(L_i))$ , otherwise, $u.hash=h(h_1||h_2||[V_\bot]||[V_\top])$, where $h_1$ and $h_2$ are value in this node's two child nodes.
\end{itemize}

 \begin{algorithm}
  \caption{\emph{Construct index merkle tree}}\label{constructe index merkle tree}
  \begin{algorithmic}[1]
  \Require The encrypted kd-kdtree $[\tau_i]$, all encrypted communication data files $L_{i1}, \dots, L_{im}$.
  \Ensure an index merkle tree $\tau_i^*$
  \State  $u= [\tau_i].rootnode$
  \If{$u$ is leafnode}\\
    \Return $u.hash = h([V_\bot]||[V_\top]||h(L_h))$, where$u.[R]=([V_\bot], [V_\top])$,$u.Pointer=P_l$ stored in leaf node in $[\tau_i]$ points to $L_h$
  \Else
    \State leaf node and right node of one node are $u_1$ and $u_2$, respectively
    \State $u.hash = h(u_1.hash||u_2.hash||[V_\bot]||[V_\top])$
    \State u = $u_i, i\in\{1, 2\}$, back to line 2
  \EndIf\\
  \Return$\tau_i^*$
  \end{algorithmic}
\end{algorithm}

\begin{figure}[H]
\centering
\includegraphics[width=3in]{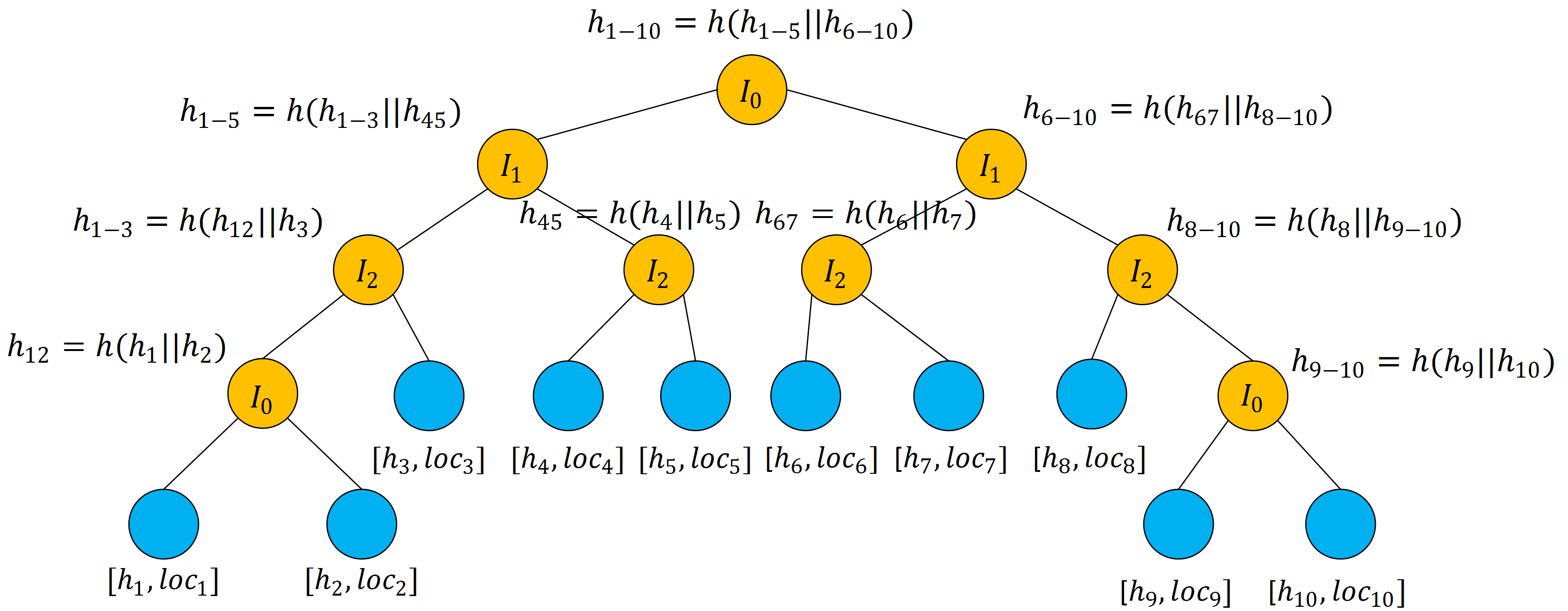}
\caption{\footnotesize{The architecture merkle tree}}
\label{index2}
\end{figure}

It is clearly that the construct of $[\tau_i]$ and $\tau_i^*$ is same according to the algorithm of constructing index merkle tree. Finally, in $\tau_i^*$, every node $u$ contains $\{ID, [R], size, Pointers, hash\}$ as shown in Fig.\ref{index2}.

Once device finishes above work, it will broadcast $L_i||Sig_c||Sig_d(h(C_i||Sig_c))$ with the corresponding index merkle tree$\tau_i^*$ to entire network.

There are many devices receiving the broadcast information. They check whether the sigature of device is legal, then mines new block. But only the miner device, which firstly finishes the predefined difficulity of PoW, can broadcast its new block into the enter network. And other device receives the new block and checks if $H(B_{i-1})$ is the right hash value of the block recently stored in block chain and if the signature of index merckle tree is valid. If both conditions are satisfied and $51\%$ of all devices approve it, the new block can be join up to main chain.

\subsection{Searching in two layers}\label{BB-Search}
In Section \ref{BB_ICDM}, steps of constructing index of two layers based on BC are presented. Now, details of searching in two layers will be described as followings. And the main goal of build this architecture is to promote accountability, i.e. user can accurately determine that the cause of system trouble is which instruction from which device and repair or change device, even though communication data file in cloud is damaged or lost.

\subsubsection{First layer:Locating the block in blockchain}
\begin{figure}[H]
\centering
\includegraphics[width=3in]{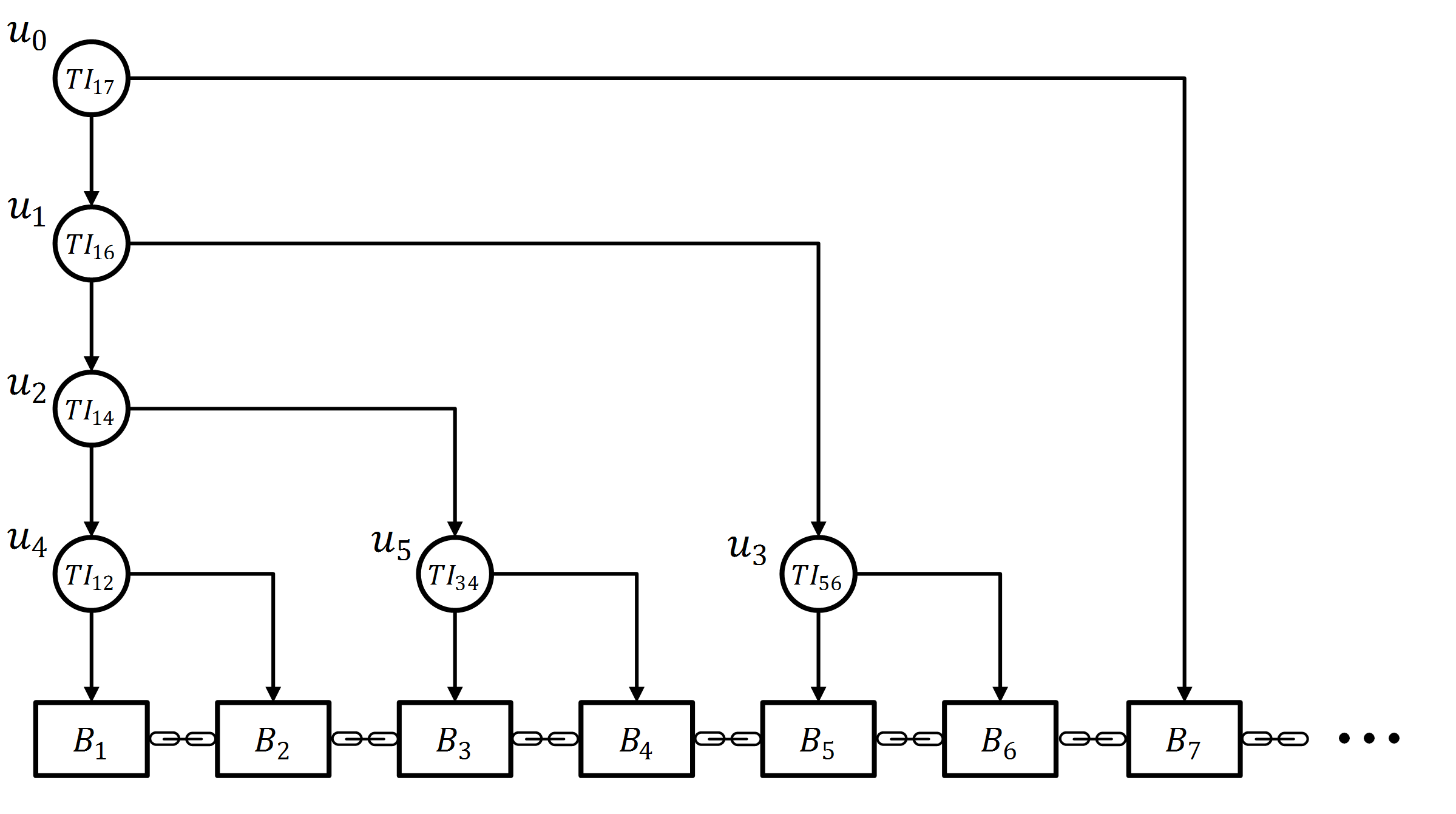}
\caption{\footnotesize{Locating the block in blockchain}}
\label{index1}
\end{figure}

First of all, user locates the block, in which the communication file he/she needs to trace. As shown in the Fig.\ref{index1}, we construsts an unbalanced index tree on block chain. Its property of unbalance can support hot and cold stratification of information, which means that newer block in BC can be easier to locate through less non-leaf node. Because of time-effectiveness of information, users pay more attention to recent block than former block in BC. In reality, user can define granularity of every leaf node, such as one week, one day even one hour. For example we use one day as the granularity in temperature control scenario mentioned before. In the index tree, every leaf node$B_i$ is generating block in $Day_i$, $i\in{1, 2, \dots, 7}$, every non-leaf node is $TI_{gh}$, $g$, $h$ is the minimum $i$ and maximum $i$ of leaf node belongs to the node. If user in that scenario finds that trouble of temperature control occured before $D_7$ , he/she download main chain from the network. And using the unbalance index tree, he/she will exclude the blocks generated after $D_6$ by check node $TI_{17}$, for the blocks gernated before $D_7$, user spends less time to checkes recent block than former block.  User checks one block by verifying whether this block stores signature signed by one of the two participants of the communication.

\subsubsection{Second layer:Searching in index merkle tree}
User searches the comminication files in the index merkle tree stored in the block that he/she located in the blockchain.
The user need to convert his/her query context to hyperrectangle $Q$, classfy $Q$'s anchor points into two part $A^{in} = (A_1^{in}, A_2^{in}, \dots, A_{2l}^{in})^T$, $A^{out} = (A_1^{out}, A_2^{out}, \dots, A_{2l}^{out})^T$, append distance information to the anchor points $A^{in}_+ = ({A^{in}}^T|(-0.5||A^{in}||^2))^T$, $A^{out}_+ = ({A^{out}}^T|(-0.5||A^{out}||^2))^T$, and encrypt $A^{in}_+$ and $A^{out}_+$ as $[A^{in}] = M^TA^{in}_+$ and $[A^{out}] = M^TA^{out}_+$ by $M$, respectively. Then user computes $r([A^{in}]-[A^{out}])\cdot[V_i]$, where $r$ is a random positive value, $[V_i]\in {[V_\bot], [V_\top]}$of hyperrectangle $[R]$, $r([A^{in}]-[A^{out}])\cdot[V_i]<0$ iff $dist(V_i, A_i^{in}) > dist(V_i, A_i^{out})$. The encrpted form of $Q$' anchor collection is trapdoor$TR=\{[A_1^{in}], [A_1^{out}]; [A_2^{in}], [A_2^{out};] \dots; [A_{2l}^{in}], [A_{2l}^{out}]\}$. Therefore the algorithm\ref{areRectsInter} combined with ASPE is shown in algorithm\ref{2areRectsInter}. Using algorithm\ref{Index merkle tree.Qry}, user can obtain a collection of leaves $L$ whose hyperrectangles intersect query hyperrectangle $Q$.

\begin{algorithm}
  \caption{\emph{RectsInter($[Q]$, $[R]$)}}\label{2areRectsInter}
  \begin{algorithmic}[1]
  \Require $[Q]$ is the encrpted form of $Q$, $Q = =\{[A_1^{in}], [A_1^{out}]; [A_2^{in}], [A_2^{out};] \dots; [A_{2l}^{in}], [A_{2l}^{out}]\}$, and $[R]$ is the encrpted form of $R$.
  \Ensure a boolean value that denotes whether $Q$ is interact with $R$
  \For{each $[A_1^{in}]$ and $[A_1^{out}]$ in $[Q]$}
    \If{$r([A^{in}]-[A^{out}])\cdot[V_\bot]<0$ and $r([A^{in}]-[A^{out}])\cdot[V_\top]<0)$}\\
      \Return false
    \EndIf
  \EndFor\\
  \Return true
  \end{algorithmic}
\end{algorithm}

\begin{algorithm}
  \caption{\emph{Index merkle tree.Qry($TR$, $\tau_i^*$)}}\label{Index merkle tree.Qry}
  \begin{algorithmic}[1]
  \Require the encryped query $TR$ and an index merkle tree $\tau_i^*$.
  \Ensure a collection of encrypted communication data files $L$
  \State $node = \tau_i^*.root$, $L=\phi$
  \If {RectsInter($u.[R]$, $TR$)}
     \If {node is a leaf node}
         \State $L=L\bigcup node$
     \Else
        \For {each $u$'s child}
          \If {RectsInter($u's child.[R]$, $TR$)}
            \State $u = u's child$, backto line2
          \EndIf
        \EndFor
     \EndIf\\
  \Return $L$
  \Else\\
    \Return $\phi$
  \EndIf
  \end{algorithmic}
\end{algorithm}

User can use pointers stored in leaf nodes in $L$ to find identifiers of communication files and download files he/she needs, if files are well preserved. But in our scheme, we provide an efficient validation method by checking encrypted communication log stored in block if files data can not be downloaded or recovered.
\section{Security performance analysis}\label{SecPerAnalysis}
\subsection{security anlysis}
We will give security anlysis in term of secure query and secure store.
\subsubsection{security anlysis on query}
\begin{itemize}\label{SAoQ}
\item\textbf {Query Privacy}: User encrpted it by ASPE approach, which has been proved to be secure in known ciphertext model, if the secret key is kept confidential[]. During download process, user find pointer in merkle tree and download the ciphertext form of $F_i$ without any leak of query.
\item\textbf {Region Privacy} In process of constructing merkle tree, ASPE approach are using to protexct information of hyperrectangle$R$. Though the whole merkle tree is exposed in block, adversary can not get anyting about region in the tree.
\item\textbf {Secure Range Query} In the search precess, ciphertext of tree protect information in tree. And tamper-resistant of blockchain is the guarantee of data completeness.
\item\textbf {Trapdoor Unlinkability}The trapdoor of query hyperrectangle
is generated based on encryption operation proposed by wang[]. And they hade prove thet the same query will be encrypted into different trapdoors.
\end{itemize}
\subsubsection{security anlysis on store}
\begin{itemize}\label{SAoS}
\item\textbf {Data Privacy} Communication data files are encrypted by utilizing the traditional symmetric key technique before submitting them to cloud. It is not within the major research scope of this paper.
\item\textbf {Non-repudiation} On one hand, user can retrive pointer of  goal communition flies on located block to prevent cloud denying. Because block in BC cannot be damaged, so if one file is record on block, user can find block records pointer.On the other hand, even if user can not retrive files from cloud, it can retrive signature of cloud and participants of communication, whivh can prevent all of three part denying.
\item\textbf {Tamper-resistant} According to the property of BC, only files that authenticated and approved by more than 51\% can join up to BC, therefore this system prevent malicious participants tampering during broadcasting process. And block chain is tamper-resistant which also protect communication log stored in exsited block from tampering.
\item\textbf {Data Retrievability} Due to the formalization of communinication logs, user can deduct approximate key information of communication data files in condition of cloud data is damaged or lost and cloud server would shift the duties onto others.
\end{itemize}

\subsection{performance anlysis}
We implement the proposed scheme using Python languagein Linux operation system and test its efficiency on a artificial data uniformly distributed of different size. All experimental results are obtained with an Intel(R) Core (TM)i5-3470 processor running at 3.74GHz, 8.00GB of RAM.

\subsubsection{Kd-tree Index Construction}
\begin{figure}[H]
\centering
\includegraphics[width=1.75in]{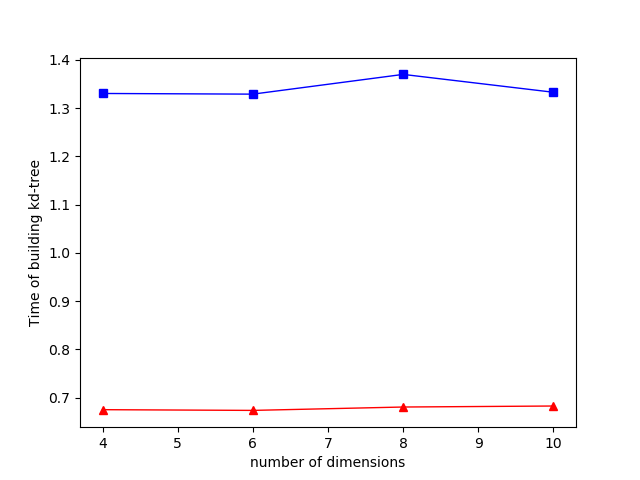}\includegraphics[width=1.75in]{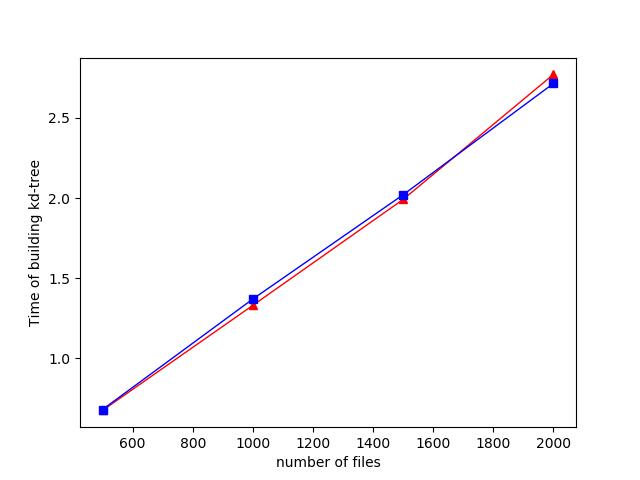}
\caption{\footnotesize{Building Kd-tree}}
\label{ki}
\end{figure}

Device builds the original kd-tree index when it wants to oursoure its communication data, which is assumed to contain $n$ $l$-dimensional file points. It will take$O(n log n)$ if an $O(n)$ median of medians algorithm, as shown in Fig\ref{ki}. The time cost of constructing kd-tree of different number of dimentions is almost changeless on the left of Fig\ref{ki}. It is increase with the increase in the number of file points in direct ratio on the right of Fig\ref{ki}.
\subsubsection{Encrypt Kd-tree}
\begin{figure}[H]
\centering
\includegraphics[width=1.75in]{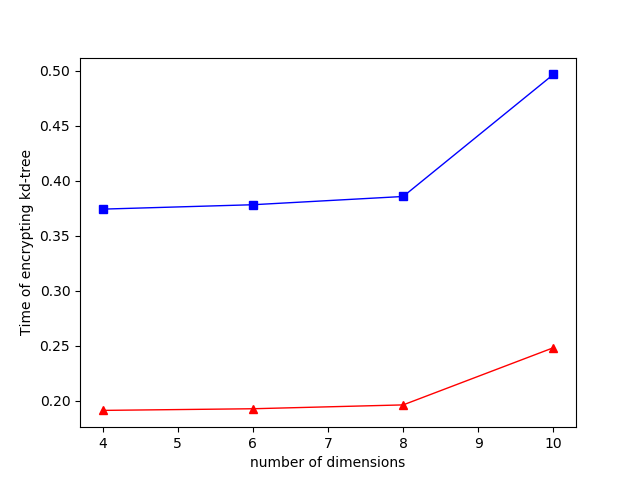}\includegraphics[width=1.75in]{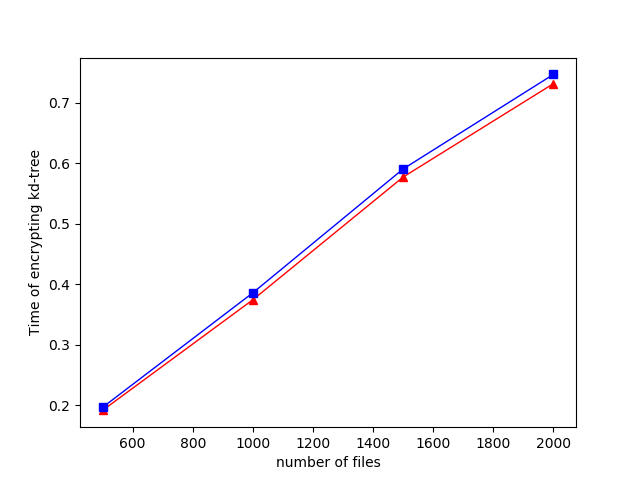}
\caption{\footnotesize{Encrypting Kd-tree}}
\label{ci}
\end{figure}

We using ASPE to encrypted kd-tree. According to the structure of kd-tree, $O(logn)$ nodes will be generated. In every node, there are two $l$-dimensional vertexs needs to be encrypted by extenting to $l$-dimensional vector and doing multiplication of a $(l+1)\times (l+1)$ matrix. Totally, the complexity of encrypting kd-tree is $O(m^2logn)$.On the left of Fig\ref{ci}, it is clearly that time cost is increasing slowly with the increase of number of dimensions, but on the right, it is clear that the time coast is proportional to points size.
\subsubsection{Merkle tree Index Construction}
After obtaining encrypted kd-tree, device building merkle tree by using  ciphertext stored in every node in kd-tree. And it is affected by points size.
\begin{figure}[H]
\centering
\includegraphics[width=1.75in]{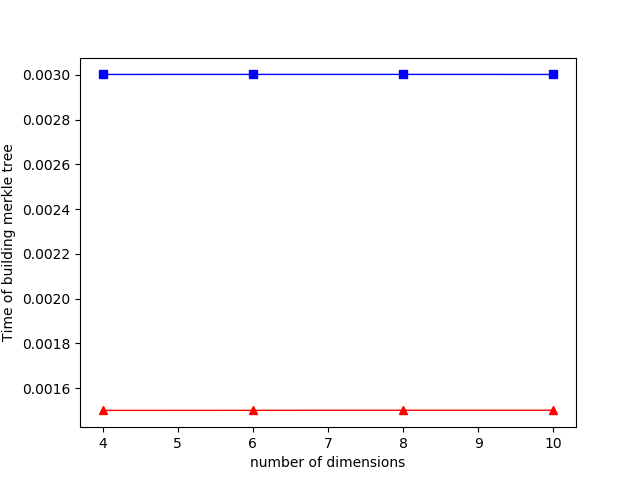}\includegraphics[width=1.75in]{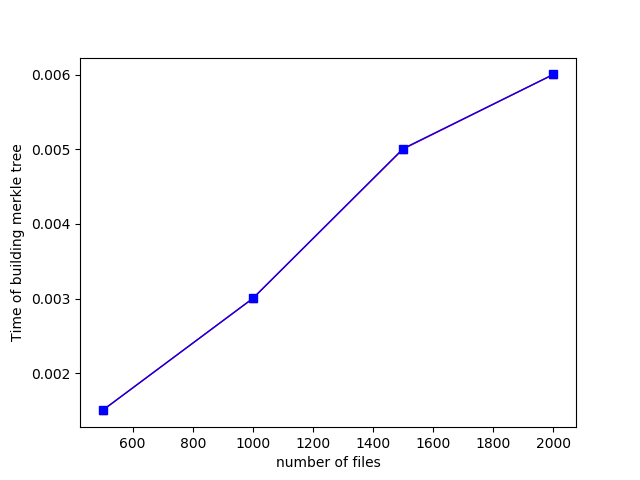}
\caption{\footnotesize{Merkle tree Index Construction}}
\label{mi}
\end{figure}
\subsubsection{Trapdoor Generation}
\begin{figure}[H]
\centering
\includegraphics[width=3in]{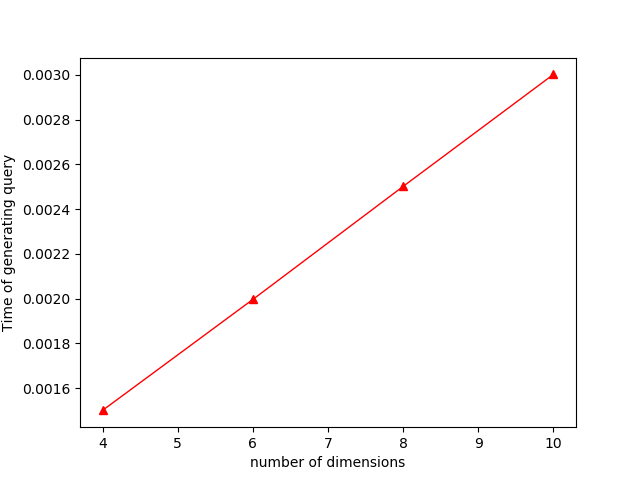}
\caption{\footnotesize{Trapdoor Generation}}
\label{qi}
\end{figure}

 User generates trapdoor from query segment by mumultipy $4l$ vectors with $(l+1)\times (l+1)$ matrix, the complexity is $O(l^2)$, and Fig\ref{qi} shows that the time cost is affected by number of dimensions.
\subsubsection{Search in First Layer}
\begin{figure}[H]
\centering
\includegraphics[width=1.75in]{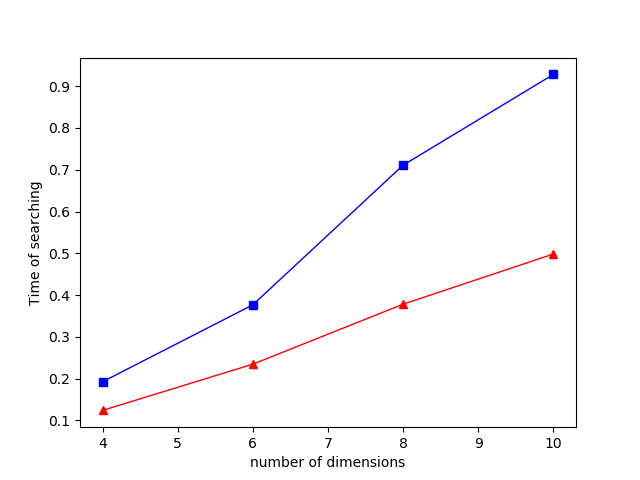}\includegraphics[width=1.75in]{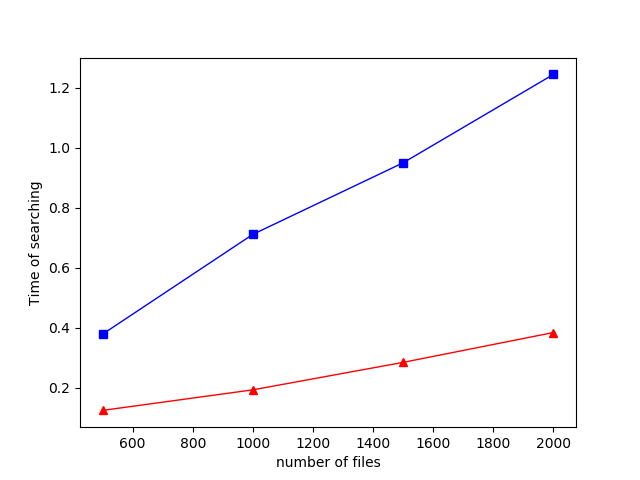}
\caption{\footnotesize{Search in First Layer}}
\label{si}
\end{figure}

One unemployed device performs search algorithm on encrypted index tree with trapdoor submited by user. Because the complexity of inner products is $O(m)$ in $m$-dimensional space, we assume that the number of leaf nodes trapdoor visited is $t$. The total number of nodes visted is $log(t)$. From the above, the whole time complexity of search is $O(mlog(t))$. Fig\ref{si} records the search time varies with different numer of dimensions and points sizes.  The bigger the points size is, the longer time needs to be used in searching process with the same number of dimensions. And with the same size of points, device needs more time to search due to increasion number of dimensions.


\section{Conclusion}\label{Conclusion}

In this paper, we firstly propose an IoT data management system based on blockchain. Each IoT devices could upload the big size communication files to cloud server with ciphertext form, but broadcast the encrypted small size communication logs to entire network, so that each IoT devices could participate to record these communication logs in a distributed database generated by blockchain technology. Under this data management model, the problem of limited data storage space, poor data reliability, poor scalability, high operating costs and poor liability judgment will be solved to some extent. For protecting privacy of both IoT devices and user, we exploit ASPE approach to encrypt the indexes of communication logs. And the new designed 2-layers index structure could provide efficient search function on blockchain database which record the IoT communication logs. Experiments on synthetic dataset demonstrate the time and space efficiency of our proposed schemes.







\ifCLASSOPTIONcompsoc
  \section*{Acknowledgments}

\else
  \section*{Acknowledgment}

\fi

This work is supported by NSFC (Grant Nos. 61502044)
\ifCLASSOPTIONcaptionsoff
  \newpage
\fi



%

\bibliographystyle{plain}
\bibliography{reference}

~~~

~~~

\noindent\footnotesize{\textbf{Hua Zhang} \textsf{Hua Zhang received B.S. degree in Communication engineering from Xidian University in 2002, M.S. degree in Cryptology from Xidian University in 2005 and PhD in Cryptology from Beijing University of Posts and Telecommunications (BUPT) in 2008. Now, she is an associate professor in Institute of Network Technology, BUPT. Her research interests include cryptography and information security. E-mail:zhanghua$\_$288@bupt.edu.cn}

~

\noindent\footnotesize{\textbf{Ziqing Guo} \textsf{
Ziqing Guo received B.S. degree in Mathematics from Beijing University of Posts and Telecommunications in 2013. Now, he is a Ph.D. candidate in Beijing University of Posts and Telecommunications. His research interests include cryptography and information security. E-mail: guoziqing@bupt.edu.cn}

~

\noindent\footnotesize{\textbf{Shaohua Zhao} \textsf{
Shaohua Zhao received B.S. degree in Mathematics and Information Science from Henan Normal University in 2015. Now, she is a Ph.D. candidate in Beijing University of Posts and Telecommunications. Her research interests include cryptography and information retrieval. E-mail: zhaoshaohua@bupt.edu.cn}

~

\noindent\footnotesize{\textbf{Qiaoyan Wen} \textsf{Qiaoyan Wen
 received the B.S. and M.S. degrees in mathematics from Shanxi Normal University, and the Ph.D. degree in cryptography from Xidian University. She is now the Leader of Network Security Center, Beijing University of Posts and Telecommunications. Her current research interests include cryptography and information security. She is a Senior Member of the Chinese Association for Cryptologic Research. E-mail:wqy@bupt.edu.cn}

\end{document}